\documentclass[twocolumn,showpacs,preprintnumbers,amsmath,amssymb]{revtex4}
\usepackage{graphicx}% Include figure files
\usepackage{dcolumn}% Align table columns on decimal point
\usepackage{bm}% bold math

\newcommand{\beq}{\begin{equation}}

\newcommand{\eeq}{\end{equation}}

\def \be  {\begin{equation}}
\def \ee  {\end{equation}}
\def \ba  {\begin{eqnarray}}
\def \ea  {\end{eqnarray}}

\begin{document}
\preprint{ITEP-TH-47/06}
\title{Non-Perturbative Decay of a Monopole:\\
the Semiclassical Pre-Exponential Factor}

\author{A. K. Monin}\email{monin@itep.ru}
\author{A. V. Zayakin}\email{zayakin@itep.ru}
\affiliation{Institute for Theoretical and Experimental Physics \\
117259, Moscow, B. Cheremushkinskaya 25, Russia } \affiliation{ M.V.
Lomonosov Moscow State University,\\ 119992, Moscow,
 Russia}

\date{\today}

\begin{abstract}The rate of the non-perturbative decay of a 't
Hooft--Polyakov monopole in an external electric field into a dyon
and a charged fermion is calculated. The sub-leading semiclassical
pre-exponential factor is presented for the first time for this
process. The leading exponential factor is shown to be in full
agreement with the previous results derived in a different
technique. Analogous treatment is shown to hold for the
two-fermionic decay of the lightest bound state in Thirring model.
Thus restoring the ``effective meson--fermion vertex'' becomes
possible.\end{abstract}

\pacs{14.80.Hv}

\maketitle

\section{Introduction}
The physics of magnetic monopoles has attracted attention for a
long time. Charge quantization~\cite{Dirac:1948um}, baryon
decay~\cite{Rubakov:1984bw},
  duality in gauge theories~\cite{Montonen:1977sn}, confinement
description~\cite{Seiberg:1994rs}
 are just a few examples of important issues associated with
 monopoles.

The 't Hooft--Polyakov monopole, which will be the main
 object of our study, is stable; its
decay is impossible unless some external field comes into play.
There exists a
 growing interest to the spontaneous and induced Schwinger
decay processes in external fields, as
 well as to the induced vacuum decay processes, see
 e.g.~\cite{Gorsky:2001up}.
Therefore it is natural to study the non-perturbatively allowed
decay of a monopole into a dyon and a fermion. This paper is
organized as follows. In Section~\ref{prelim} some general facts
on monopole physics and induced decays are reviewed. We examine
the conditions under which a semiclassical treatment is valid for
the considered problem. The decay rate is calculated in
Section~\ref{preexp}. The elaborated technique is simplified and
applied to the bound state decay of the Thirring model in
Section~\ref{thirmod}, and the results are summarized in
Section~\ref{final}.

\section{\label{prelim} Preliminaries}
\subsection{Monopoles: Non-Perturbative and Non-Local
Objects} Since the historic paper by Dirac~\cite{Dirac:1948um} the
question how to incorporate the dynamics of magnetic monopoles into
the standard quantum field-theoretical paradigm has been
non-trivial. Treating monopoles and charges within the same
framework is hindered by the two obstructions: inapplicability of
the perturbation theory and non-locality.

Due to Dirac's quantization condition~\cite{Dirac:1931kp}, the
charge $g$ of a monopole is $g^2\sim \frac{1}{\alpha}\sim 137$ so
that no reasonable perturbation series can be derived with respect
to this parameter, unlike the standard QED perturbation theory in
powers of $\alpha$. Several attempts have been made to elaborate a
self-contained QED with monopoles \cite{Zwanziger:1970hk,
Gamberg:1999hq}.

These two fundamental problems inevitable for the point-like Dirac
monopoles arise under a different guise for the 't~Hooft--Polyakov
monopole. The monopole configuration is {\it a priori} a solution
to the classical field equations. It exhibits some properties of a
point particle, but it cannot be treated as if it were generated
by some local field operator\footnote{In simpler cases, e.g. in
sine-Gordon theory, a quantum solitonic object may be written down
as an explicitly given non-local field operator (the so-called
Mandelstam operator~\cite{Rajaraman:1982is}). Monopole creation
operator is known in lattice gauge
theory~\cite{Frohlich:1998wq}.}. The 't Hooft--Polyakov monopole
should be thought of as a kind of semiclassical object rather than
a quantum particle since its characteristic size is roughly
$1/\alpha$ times greater than its de~Broglie wavelength.  In the
dual theory~\cite{Montonen:1977sn} the monopoles correspond to the
original gauge bosons, which do have a local description; however,
in the original theory itself no local description is possible.

The non-perturbative issues of monopole dynamics can be studied
via geometric and topological methods, permitting description of
dynamics of monopoles and \hbox{dyons} in terms of geodesics on
the moduli spaces of solutions to the Bogomolny
equations~\cite{Atiyah:1985fd,Gibbons:1986df}. Processes which
have an explicit quantum field-theoretical {\it interpretation}
like scattering of monopoles into monopoles or dyons have been
shown to take place. However, no quantum field theoretical {\it
model} of these processes exists so far.

String theory suggests  describing  dyons as $(p,q)$ strings with
ends fixed on some D-branes~\cite{Polchinski:1998rq}. This
description was recently proposed to induce the process ``gauge
boson $\to$ monopole, dyon'' or ``monopole $\to$ dyon, charge'' in
an external field~ by Gorsky, Saraikin and Selivanov
\cite{Gorsky:2001up}. The existence of a corresponding vertex in
string theory is mentioned to show that there are some attempts to
elaborate a local perturbative-like treatment of monopoles. The
string vertex is not directly used
 in the calculations below. However, its existence
provides us with a heuristic apology to validate introducing  an
``effective coupling'', which is absent at the perturbation theory
level.

There exists a wide class of processes in field theory becoming
non-perturbatively allowed once an external field comes into play.
The obvious example is the Schwinger spontaneous $e^+e^-$ pair
production in an external electric field (for a review
see~\cite{Dunne:2004nc}), or an analogous process for the
spontaneous Schwinger-like monopole pair production in the static
magnetic field~\cite{Affleck:1981ag}. Another class of
non-perturbative phenomena consists of false vacuum decay processes
in a scalar field theory. The generic case of false vacuum decay in
a distorted Higgs-like potential was initially discussed
in~\cite{Kobzarev:1974cp,Coleman:1977py}. There exists a deep
similarity between spontaneous Schwinger processes and false vacuum
decay. Formally, these two phenomena are identical in $1+1$
dimensions~\cite{Voloshin:1985id}. The action $S_{cl}$ of a
classical configuration of $e^+e^-$ paths in Euclidean domain
contributing to the semiclassical pair creation  probability $w\sim
e^{-S_{cl}}$ behaves like
``$const_1\mathrm{(volume)}-const_2\mathrm{(surface)}$''. The same
behaviour is typical for the action of a classical bubble in the
thin wall approximation, describing, in its turn, a semiclassical
vacuum decay probability. This statement can be considered as a hint
to a better understanding of more general cases for the both types
of processes.

\subsection{Induced vs. Spontaneous}

History of the false vacuum decay teaches us a lesson that if a
process is possible as a spontaneous one, there should exist
related induced ones~\cite{Affleck:1979px,Selivanov:1985vt}. The
same argument works for the Schwinger processes. A possibility of
an induced Schwinger-type monopole decay was first suggested
in~\cite{Gorsky:2001up}. Monopoles were first treated as triggers
for vacuum decay in a scalar field theory long
ago~\cite{Steinhardt:1981ec}.

This interpretation allows one to symbolically introduce an
 effective
``charge--monopole--dyon'' vertex, although it does not exist at
the level of perturbation theory. As in our previous
papers~\cite{Monin:2005wz,Monin:2006dt}, 't Hooft--Polyakov
monopole is treated as a semi-classical object, for which the
notion of the trajectory is well defined. Only trajectories far
larger than the monopole size are dealt with, in order not to
break down the semiclassical approximation. The trajectory of the
monopole is analytically continued into the Euclidean domain,
where a correction to its Green function is calculated, yielding
the decay rate.

We can not describe monopole in terms of a second-quantized theory.
What is meant here then by ``the Green function of a monopole''?
This Green function stands for an effective one-particle
description. One is incapable of writing down a quantum field
theoretical path integral for it, nevertheless, a 1-particle
quantum-mechanical path integral
 for a particle
with a given spin, electric charge $e$ and magnetic charge $g$ in an
external vector-potential $A_\mu$ is meaningful in the semiclassical
approximation.

The close relation of the present problem to the issue of false
vacuum induced decay has already been pointed out. In the course
of calculations, both problems are dealt in a semiclassical
technique very close to that of world-line instantons by Dunne and
Schubert~\cite{Dunne:2005sx}. Therefore, the structure of the
result is similar:
$$\Gamma\sim Ke^{-S_{cl}},$$ \noindent where the leading exponent
behavior  is governed by  the action on a classical configuration
$S_{cl}$, be it a field distribution in field theory or a
1-particle trajectory in quantum mechanics; the subleading
pre-exponential factor $K$ generally costs more efforts to be
extracted~\cite{Dunne:2006st}. It contains the fluctuation
determinants as well as contributions from Jacobians, which arise
when integrating out the collective coordinates.

Basically, two techniques exist for calculating this prefactor. One
can either study the fluctuation determinant of the operator
describing oscillations around the classical
solutions~\cite{Kiselev:1975eq}
 or one can reduce the
field-theoretical problem to that of 1-particle relativistic
quantum mechanics and obtain the prefactor in terms of the WKB
method~\cite{Voloshin:1985id}.

The level of complexity of the prefactor calculation depends on
the method applied. E.g., the prefactor in Schwinger's derivation
of $e^+e^-$ production rate comes at the same price with the
exponent. On the other hand, when time-dependent field enters the
play it often comes out to be useful to  calculate the
determinants via the Gelfand--Yaglom or
Levit--Smilansky~\cite{Levit:1976fv} method, or via the Riccati
equation method~\cite{Kleinert:2004ev}.

In a paper by one of us (A.K.M.)~\cite{Monin:2005wz} the monopole
decay was studied by means of Feynman path integrals in the
leading semiclassical approximation. Proof of the existence of a
negative mode in the spectrum was also given, however, the full
fluctuation determinant was not calculated. In our preceding paper
this technique was extended to inhomogeneous
fields~\cite{Monin:2006dt}. Here  a calculation giving the
exponential and the pre-exponential factor simultaneously is
presented.

\section{\label{preexp}Monopole in 4D}
A monopole with a magnetic charge $g$, mass $M_m\sim M_W/\alpha$
($M_W$ is the $W$-boson mass) is considered in a constant external
electric field in a four-dimensional space-time. The rate of its
decay into a dyon of mass $M_d$ with electric and magnetic charges
$e,g$ respectively, and a charged fermion of mass $m_e$ will be
calculated. First the reader is reminded how Green functions can
be obtained for an electrically and magnetically charged particle
in an external field. Then  a ``loop correction'' is calculated,
although this notion has a limited applicability, as commented
above.

It has already  been mentioned that the monopole Green function has
got only a semiclassical meaning in the proposed approach. This
means that one is bound by the requirement for the charge-dyon loop
to be larger than the 't~Hooft--Polyakov monopole size. Technically
this will imply taking all loop integrals in the saddle-point
approximation. On the other hand, the saddle--point approximation
does a good job: it yields the imaginary part of mass correction
directly, avoiding the infinite real mass renormalization
part~\footnote{Monopole mass renormalization due to quantum
fluctuations over the classical configuration was discussed
in~\cite{Kiselev:1988gf}. Mass correction was found out to contain
quadratic and logarithmic divergences. After renormalization, finite
real non-perturbative mass correction $\delta
M_m=-\frac{M_W}{2\pi}\log\frac{M_W}{M_H}$ was found, where $M_W,M_H$
are the $W$-boson and the Higgs masses respectively. Here mass
correction due to a different effect is calculated, namely, induced
Schwinger process, not considered in~\cite{Kiselev:1988gf}. However,
an infinite part of mass correction is implicitly present in our
calculation through divergences at $\alpha_i=0$ in the
expression~(\ref{contr}) below, avoided by taking the saddle-point
approximation}.

A self-consistent field-theoretical treatment of Abelian monopoles
not requiring introduction of Dirac strings was performed by
Zwanziger. Let us consider fermionic fields $\psi_i$ carrying both
electric charge $e_i$ and magnetic charges $g_i$. Then two $U(1)$
currents will be describing the interaction of the system with the
external field, electric current $j_e$ and magnetic current $j_g$

\beq j_e^\mu=\sum_i e_i \bar{\psi}_i\gamma^\mu \psi_i\eeq \beq
j_g^\mu=\sum_i g_i \bar{\psi}_i\gamma^\mu \psi_i\eeq

which are subject to condition

\beq
\begin{array}{l}
F^{\mu\nu}=\partial^\mu A^\nu-\partial^\nu A^\mu +(\mbox{non-local
terms}) \\

\tilde{F}^{\mu\nu}=\partial^\mu\tilde{A}^\nu-\partial^\nu
\tilde{A}^\mu +(\mbox{non-local terms})
\end{array}
\eeq

Interaction Lagrangian will then be organized as

\beq L_{int}=j_\mu^e A^\mu+j_\mu^g \tilde{A}^\mu \eeq

It is argued by Zwanziger~\cite{Zwanziger:1970hk} that the
non-local terms have zero-measure support and thus they can be
neglected for practical purposes. Moreover, in present case the
non-local terms may be neglected due to the non-Abelian nature of
the initial field configuration.

The Green function for a scalar particle with electric charge $e$
and magnetic charge $g$ can be given in terms of first-quantized
formalism suggested by Affleck et al.~\cite{Affleck:1981bm}

$$G(y,x)=\int ds e^{im^2s}\int^{\tiny\lefteqn{x(s)=y}}_{\tiny\lefteqn{x(0)=x}}\mathcal{D}x(t) e^{i\int_0^s
\dot{x}^2/4+e\int A_\mu dx^\mu + g\int\tilde{A}_\mu dx^\mu}$$

In a constant external field this can be calculated exactly. On the
other hand, this Green function is nothing else than the matrix
element
$$G(y,x)=\left\langle
y\left|\frac{1}{D^2+m^2}\right|x\right\rangle.$$ The covariant
derivative for a particle with both electric and magnetic charges
$e$ and $g$ in an external field should look like
\begin{equation}\nonumber
D_\mu=\partial_\mu+ieA_\mu+ig\tilde{A}_\mu.
\end{equation}
 This was justified by Gibbons and
Manton~\cite{Gibbons:1995yw}.

First-quantized treatment exists for fermions as well, but it is
easier for us to write down the fermionic Green function by virtue
of similarity

$$G_{F}(y,x)=\left\langle
y\left|\frac{1}{m-i\hat{D}}\right|x\right\rangle$$

Consider now a constant electric field ${\bf E}=(0,0,E)$. Let us
choose a vector potential in the form
$A_\mu(x)=\frac{E}{2}(-x_3,0,0,x_0)$, hence
$\tilde{A}_\mu=\frac{E}{2}(0,-x_2,x_1,0)$. The Dirac operator takes
the form
\begin{equation}\label{Dirac operator}
i\hat{D}-m=i\gamma^\mu
D_\mu-m=i\gamma^\mu(\partial_\mu+ieA_\mu+ig\tilde{A}_\mu)-m.
\end{equation}
A propagator of a fermionic particle with an electric charge $e$ and
magnetic charge $g$ is given by
\begin{equation}\label{euclgr}
G_{F}(y,x)=(m+i\hat { D_y } )G^{(0)}(y,x),
\end{equation}
where the following auxiliary function is introduced
\begin{equation}\begin{array}{lll}
\displaystyle G^{(0)}(y,x)&=&-\displaystyle\frac{i}{32\pi^2}\,egE^2 \times \\
&\times\displaystyle\int\limits_0^{\infty}&\!\displaystyle
ds\frac{\mathrm{e}^{i\frac{(m^2+i\epsilon)s}{2}}\mathrm{e}^{-\frac{1}{2}(
eEs\gamma^0\gamma^3+gEs\gamma^1\gamma^2)}\mathrm{e}^{iS}}
{\sinh\left(\frac{eEs}{2}\right)\sin\left(\frac{gEs}{2}\right)}
\label{G_0}
\end{array}
\end{equation}
Terms $i\epsilon$ will be omitted further. Here
\begin{equation}\begin{array}{lll}
S=\frac{eE}{4}(y-x)_{\|}^2\coth\frac{eEs}{2}+\frac{eE}{2}
(y_0x_3-y_3x_0) +\\
+\frac{gE}{4}(y-x)_\bot^2\cot\frac{gEs}{2}+\frac{gE}{2}
(y_1x_2-y_2x_1),\end{array}
\end{equation}
indices $\|$ and $\bot$ denote the $(0,3)$ and $(1,2)$ components of
4-vector correspondingly. Deforming the $s$ integration contour
(roughly speaking, turning it like $s\to is$)\footnote{As can be
seen from (\ref{G_0}), the integrand contains term like
$\exp(\coth(z))$, possessing essential singularities at $z=\pi in$.
Therefore this transformation is not a pure rotation $s\to is$ but
rather a deformation which must avoid traversing the singularity
points.} and making a transition to Euclidean quantities like
$x_0\to -ix_0$, one writes down the Euclidean Green function
\begin{equation}\begin{array}{lll}
G_E^{(0)}(y,x)&=&\displaystyle\frac{1}{32\pi^2}\,egE^2\times \\
&\times\displaystyle\int\limits_0^{\infty}&\!\displaystyle
ds\frac{\mathrm{e}^{-\frac{m^2
s}{2}}\mathrm{e}^{\frac{1}{2}(eEs\gamma^0\gamma^3
+igEs\gamma^1\gamma^2)}\mathrm{e}^{-S_s}}{\sin(\frac{eEs}{2})\sinh(\frac{gEs}{2})}\,,
\label{G_0_E}
\end{array}
\end{equation}
where
\begin{equation}\nonumber\begin{array}{l}
S_s=\frac{eE}{4}(y-x)_{\|}^2\cot\frac{eEs}{2}-\frac{eE}{2}
(y_0x_3-y_3x_0)+\\
+\frac{gE}{4}(y-x)_\bot^2\coth\frac{gEs}{2}-i\frac{gE}{2}
(y_1x_2-y_2x_1),\end{array} \label{action_E}
\end{equation}
all the four-vectors in this expression are supposed to be taken in
Euclidean space with the positive overall metric sign; the index $E$
will be omitted further. The fermionic propagator thus takes the
form
\begin{equation}\label{fermion propagator}
G_F(y,x)=\left(m+\gamma^\mu a_\mu(y,x)\right)G^{(0)}(y,x),
\end{equation}
where
\begin{equation}
\begin{array}{lll}
a_\|(y,x)&=&\Bigl(\frac{eE}{2}(y_0-x_0)\cot\alpha+\frac{eE}{2}(y_3-x_3),
\\&&\qquad\frac{eE}{2}(y_3-x_3)\cot\alpha-\frac{eE}{2}(y_0-x_0)\Bigr), \\
a_\bot(y,x)&=&\Bigl(\frac{gE}{2}(y_1-x_1)\coth\beta+i\frac{gE}{2}(y_2-x_2),
\\&&\qquad\frac{gE}{2}(y_2-x_2)\coth\beta-i\frac{gE}{2}(y_1-x_1)\Bigr),
\end{array}
\end{equation}
with $\alpha=\frac{eEs}{2}$ and $\beta=\frac{gEs}{2}$.

There are arguments in favour of thinking (0,1)-monopole to be a
scalar particle and a (1,1)-dyon to be a spin-$\frac{1}{2}$
particle~\cite{Coleman:1982cx}, thus fermionic Green function
above refers to dyons.  It describes charged fermions as well in
the limit $g=0$.

The correction to monopole's Green function propagating from
$(0,0,0,0)$ to $T=(0,0,0,T)$ may be expressed in terms of Feynman
path integrals~\cite{Monin:2005wz} and reduced to a contraction of
Green functions (here the ``effective vertex'' of
monopole-dyon-charged fermion interaction is suggested to be of
the form $\,\lambda\phi\bar{\psi}\psi$)
\begin{equation}\begin{array}{lll}\label{contr}\delta
G_m(T,0)&=&\displaystyle\lambda^2\int G_m(z,0)G_m(T,w)\times\\
&&\qquad\displaystyle\times\:\mathrm{tr}[G_e(w,z)G_d(w,z)]dw\,dz,
\end{array}
\end{equation}
$\lambda$ being (an unknown\footnote{In the next section  some
arguments will be given for restoring $\lambda$ form a different
calculation                               in the
2-dimensional case.}) dimensionless
factor, indices $m,e,d$ belonging here and everywhere below
to a
monopole, a charged fermion, and a dyon respectively.
Substituting the above Green functions  for their Schwinger
representations~(\ref{G_0}), one can express the trace
in~(\ref{contr}) in terms of Schwinger parameters $\alpha_i$
\begin{equation}\label{Trace}\begin{array}{lll}
\mathrm{Tr}&\equiv&\mathrm{tr}(M_d+\hat{a}
)(\cos\alpha_2+\gamma^0\gamma^3\sin\alpha_2)\times \\ &\times&
(\cosh\beta_2+i\gamma^1\gamma^2\sinh\beta_2)\times \\ &\times&
(m_e+\hat{b})(\cos\alpha_1-\gamma^0\gamma^3\sin\alpha_1),\end{array}
\end{equation}
here $a$, $\alpha_2$, $\beta_2\equiv\frac{g}{e}\alpha_2$ correspond
to the  dyon propagator and $b$, $\alpha_1$ to that of the charged
fermion. Schwinger parameters $\alpha_3$, $\alpha_4$
coresponding to monopole propagation are also present in~(\ref{contr}).
Calculating the trace one obtains
\begin{equation}\nonumber
\label{calculated trace}
\begin{array}{lll}
\mathrm{Tr}&=&\displaystyle4\Biggl((m_eM_d\cosh(\frac{g}{e}
\alpha_2)\cos(\alpha_1-\alpha_2)+\\&+&\displaystyle\left(\frac{eE}{2}\right)^2
(w-z)_\|^2\frac{\cosh(\frac{g}{e})\alpha_2}{\sin\alpha_1\,
\sin\alpha_2}+ \\ &+&\displaystyle \frac{egE^2}{4}
(w-z)_\bot^2\frac{\cos(\alpha_1-\alpha_2)}{\sinh(\frac{g}{e}
\alpha_2)}\Biggr).
\end{array}
\end{equation}
Performing Gaussian integrals over $z$ and $w$, and
introducing Feynman variables
$\alpha_3=Ax,\alpha_4=A(1-x),$ with the Jacobian of the substitution
being $A$, one notes that no dependence on $x$ enters
formula~(\ref{contr}), thus the $x$-integration is taken off
trivially, after which the correction to Green function becomes
\begin{widetext}
\begin{equation}\label{delta G befor int A}
\begin{array}{lll}
\delta G &=&const\displaystyle\int\frac{d\alpha_1\,d\alpha_2\,A\,dA
}
{\alpha_1\,\sin\alpha_1\,\sin\alpha_2\,\sinh(\frac{g}{e}\alpha_2)}\displaystyle\frac{\displaystyle
\mathrm{\displaystyle
e}^{-\left[\frac{m_e^2}{eE}\alpha_1+\frac{M_d^2}
{eE}\alpha_2+\frac{M_m^2}{eE}A+
\frac{\frac{eE}{4}T^2}{A+\frac{\sin\alpha_1\,\sin\alpha_2}{
\sin(\alpha_1+\alpha_2)}}\right]}}{\displaystyle\Bigl[\bigl(\frac{e}{\alpha_1}+g\cot\frac{
g\alpha_2}{e}\bigr)\sinh\frac{gA}{e}
+g\cosh\frac{gA}{e}
\Bigr]\displaystyle\bigl[A(\cot\alpha_1+\cot\alpha_2)+1\bigr]}\times
\\
&\times&\displaystyle\Biggl\{m_eM_d\cosh(\frac{g}{e}
\alpha_2)\cos(\alpha_1-\alpha_2)+\displaystyle eE\frac{\cosh(\frac{g}{e}
\alpha_2)A}
{\sin\alpha_1\,\sin\alpha_2[A(\cot\alpha_1+\cot\alpha_2)+1]}
+
\\
&+&\displaystyle\Bigl(\frac{eET}{2}\Bigr)^2\frac{\cosh(\frac
{ g}{e}\alpha_2)}
{\sin\alpha_1\,\sin\alpha_2[
A(\cot\alpha_1+\cot\alpha_2)+1]^2 }+\displaystyle
\frac{egE\cos(\alpha_1-\alpha_2)\,\sinh(\frac{g}{e}A)}{\alpha_1
\sinh(\frac{g}{e}\alpha_2)\displaystyle\Bigl[\bigl(\frac{e}{\alpha_1}
+g\cot\frac{g\alpha_2}{e}\bigr)\sinh\frac{gA}{e}
+g\cosh\frac{gA}{e}\bigr)\Bigr]}\Biggr\}.
\end{array}
\end{equation}
\end{widetext}
To integrate over variable $A$, the saddle-point approximation is
employed. Generally, the saddle-point approximation works for the
integrals \beq\int_0^{+\infty}e^{\nu
f(s)}g(s)ds=\sqrt{\frac{2\pi}{-\nu f''(s_0)}}e^{\nu
f(s_0)}g(s_0)+\mathrm{O}\left(\frac{1}{\nu}\right)\eeq [$s_0$
being the minimum point of $f(s)$], when $\nu\to\infty$. In the
present case, \beq \nu f(A)=-\frac{M_m^2}{eE}\left[A+
\frac{(eE)^2}{4M_m^2}T^2\frac{1}{A+const}\right]\eeq satisfies
this requirement since  $\nu=\frac{M_m^2}{eE}$ is a large
parameter indeed, coefficient $\frac{(eE)^2}{4M_m^2}T^2$ being not
infinitesimal as $T$ may be made large enough for our purposes. In
fact, the limit $T\to\infty$ will be used, so the latter statement
is fairly justified.

\begin{figure}[t]
\begin{center}
\includegraphics[height = 5cm, width=2.5cm]{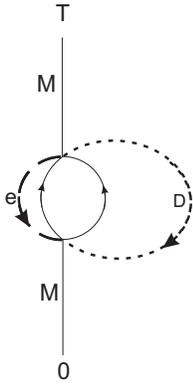}
\caption{\label{Instantons} Classical paths in $(x_3,x_0)$
plane
with arbitrary winding numbers. }
\end{center}
\end{figure}

The saddle point value $A_0$ in the integral~(\ref{delta G befor int
A}) over $A$  is assumed to satisfy $A_0\gg 1$, so in principle one
could consider asymptotics for hyperbolic functions in the form
$\sinh\frac{gA}{e}\approx\cosh\frac{gA}{e}\approx\frac{1}{2}
\,\mathrm{e}^{\frac{gA}{e}}$, and raise $\frac{gA}{e}$ to
 the exponent. However, one should remember that since
 the monopole and the dyon are
being treated as point-like particles, it is obligatory to consider
an external field small enough so that the size of the loop (see
Fig.~\ref{Instantons}) is larger than the size of the monopole.

For such a field it is easy to show that $\frac{M^2_m}{gE}\gg
g^2$. So the term $gA/e$ must be neglected in the exponent
compared to $m^2A/eE$. But $gA/e$ is still large enough to
consider hyperbolic functions $\cosh(\frac{g}{e}A)$ and
$\sinh(\frac{g}{e}A)$ approximately equal. Then the saddle point
value for $A$ is
\begin{equation}\nonumber
A_0=\frac{eET}{2M_m}-\frac{\sin\alpha_1\/\sin\alpha_2}{
\sin(\alpha_1+\alpha_2)},
\end{equation}
and the second derivative is
\begin{equation}\nonumber
\frac{\partial^2f}{\partial A^2}=\frac{4 M_m^3}{(eE)^2T}.
\end{equation}

In order to find the monopole mass correction one should know the
asymptotic form of the propagator of a scalar particle in an
external field. The scalar Euclidean  propagator has the following
asymptotics
\begin{equation}\label{monopole propagator}
G_m(T,0)=\frac{1}{16\pi^{3/2}}\frac{gE}{\sqrt{M_mT}}\,
\frac { \mathrm { e } ^ { -M_mT }
}{\sinh\frac{gET}{2M_m}} ,
\end{equation}
and the leading-order (in powers of $T$) contribution to
its
variation due to the variation of the monopole mass
\begin{equation}\label{variation of monopole propagator}
\delta G_m(T,0)=-\frac{1}{8\sqrt{2}\pi^{3/2}}\delta
M_m\,gE\sqrt{\frac{T}{M_m}}\,\frac{\mathrm{e}^{-M_mT}}{\sinh
\frac{gET}{2M_m}} .
\end{equation}
Comparing this result with the one obtained after
integration
(\ref{delta G befor int A}) over $A$ one gets the
mass
correction
\begin{equation}\label{22}
\begin{array}{lll}
\delta
M_m&=&\displaystyle\frac{\lambda^2
g}{(32\pi)^{3/2}M}\displaystyle\int\frac{d\alpha_1\,d\alpha_2}
{\alpha_1\,\sinh(\frac{g}{e}\alpha_2)\,\sin(\alpha_1+\alpha_2)}
\times \vspace{0.4cm}\\ &\times&\frac{\displaystyle\mathrm{\displaystyle
e}^{\displaystyle-\left(\frac{m_e^2}{eE}\alpha_1+\frac{M_d^2}{eE}
\alpha_2-\frac{M_m^2}{eE}
\frac{\sin\alpha_1\,\sin\alpha_2}{\sin(\alpha_1+\alpha_2)}\right)}\vspace{0.2cm}}
{\displaystyle\bigl(\frac{e}{\alpha_1}+g\cot(\frac{g}{e}\alpha_2)+g\bigr)}\times\\
&\times&\displaystyle
\biggl[m_eM_d\cosh\left(\frac{g\alpha_2}{e}\right)
\cos(\alpha_1-\alpha_2)
+\\&&\displaystyle\qquad +M_m^2\cosh\left(\frac{g\alpha_2}{e}\right)\frac{
\sin\alpha_1\sin\alpha_2}{\sin^2(\alpha_1+\alpha_2)}\biggr].
\end{array}
\end{equation}
The terms proportional to $E$ compared to the ones proportional to
any bilinear combination of masses have already been neglected here.
It was reasonable to leave them out since such an assumption had
already been taken when integrating over $A$ in the saddle-point
technique.
The last step is to integrate over $\alpha_1$ and $\alpha_2$ using
the saddle point method. Note that the custom integration via
methods of the theory of complex variable functions fails,
due to an
essential non-analyticity in $\alpha_1+\alpha_2$ present in the
expression being studied (roughly speaking, it is like $e^{-1/z}$ in
the vicinity of $z=0$, as can be seen from~(\ref{22}) above). On the
contrary, saddle-point approximation remains valid, because all
massive parameters are considered to be large compared to
$\sqrt{eE}$.

\begin{figure}[t]
\begin{center}
\includegraphics[height = 5cm, width=8cm]{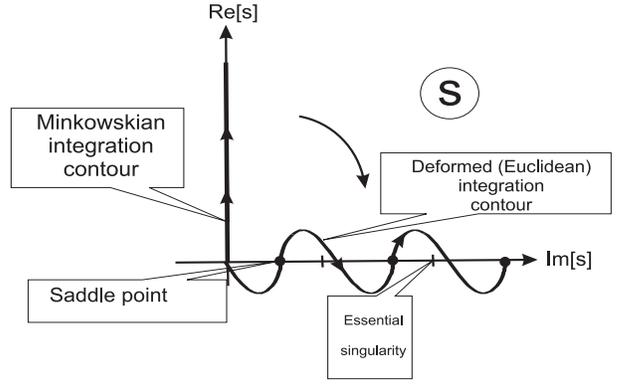}
\caption{\label{contour}Integration contour for Minkowskian and
Euclidean Green Functions.}
\end{center}
\end{figure}

However, due to the specified essential singularities, a complicated
deformation of the integration  contour should be performed.
Formula~(\ref{22}) should rather be understood in the following way:
one starts with the Minkowskian Green functions, for which path of
integration is directed along the imaginary axis of $z\equiv
\alpha_1+\alpha_2$, being away from essential singularities. Such a
contour rotation refers not only to~(\ref{22}), but
to~(\ref{euclgr}) and (\ref{G_0}) as well. The original Minkowskian
Green function was defined with a contour directed along imaginary
$s$ axis. When writing down the Euclidean Green
function~(\ref{euclgr}), one should already have given a
prescription for turning the integration contour to the real $s$
axis. How it should have been done is shown in Fig.~\ref{contour}.
Here singularities do not lie on integration path; and
saddle-points are passed in the (imaginary) direction prescribed
by steepest descent condition. The deformation was performed in
the domain of analyticity of the integrand, without traversing the
singularities. The integral is dominated by saddle points, and may
be evaluated as sum of integrals in the vicinities of each
saddle-point. A contour (of real dimension 2) in $\mathbb{C}^2$
for~(\ref{22}) is constructed in a similar way. It is not shown
here due to high dimensionality.

The function
$f(\alpha_1\,,\alpha_2)=\frac{m_e^2}{eE}\alpha_1+\frac{M_d^2 }{eE}
\alpha_2-\frac{M_m^2}{eE}
\frac{\sin\alpha_1\,\sin\alpha_2}{\sin(\alpha_1+\alpha_2)}$ is to be
minimized. One gets the saddle point values $\theta_i^{(n)}$ for
$\alpha_i$, which come out to be the same
as were obtained
in~\cite{Monin:2005wz} by a different method
\begin{equation}\nonumber
\begin{array}{lllllll}
\left(\begin{array}{c}
\theta_1^{\pm(n)}\\
\theta_2^{\pm(m)}\end{array}\right) &=& \pm\left(\begin{array}{c}
\cos^{-1}\frac{M_m^2+m_e^2-M_d^2}{ 2m_eM_m }\\
\cos^{-1}\frac{M_m^2-m_e^2+M_d^2}{ 2M_dM_m}
\end{array}\right)+\left(\begin{array}{c}
2\pi n \\ 2\pi m \end{array}\right)
\equiv \\ &\equiv&\left(\begin{array}{c}
\pm \theta_1+2\pi n \\ \pm\theta_2+2\pi m
\end{array}\right),\\
&&n,\, m\in\mathbb{Z},\theta_i^{\pm(n)}>0

\end{array}
\end{equation}
the corresponding
determinant being
\begin{equation}
\begin{array}{lll}
\displaystyle\det\vphantom{|}_{ij}\left(\frac{\partial^2f}{
\partial\alpha_i\partial\alpha_j }
\right)&=&\displaystyle -4\frac{\sin^2\theta_1\,\sin^2\theta_2}
{\sin^4(\theta_1+\theta_2)}\left(\frac{M_m^2}{eE}
\right)^2= \\ &=&\displaystyle -4\frac{(m_e M_d)^2}{(eE)^2}.
\end{array}
\end{equation}
One can see that there exists a two-parameter family of local minima
of the saddle-point integral. Geometrically, the integer parameters
$m,n$ denote multiply-wound classical solutions. The result is a sum
over all saddle points. The physical meaning of such a sum was
discussed in~\cite{Monin:2006dt}. The semiclassical approximation
counts all possible classical sub-barrier trajectories, which are
arcs of a circle, $\theta_i$ having direct meaning of an angular
coordinate on the particle trajectory in the Euclidean plane, taking
them with weights $e^{-S^{\pm}_{n,m}}$ given below.

Finally one obtains the mass correction as a sum over
winding numbers $m,n$
\begin{widetext}
\begin{equation}\nonumber
\begin{array}{lll}
\displaystyle\mathrm{Im}\,\delta
M_m&=&\displaystyle-\frac{\lambda^2}{8\pi}\,
\displaystyle\frac{ eE}{M_m}\displaystyle\left\{\displaystyle\sum_{n,m=0} \frac{\mathrm
{ e } ^ { -S^{+}_{n,m}}\cos^2(\frac{\theta_1 -\theta_2 } { 2
} ) } {\sin(\theta_1+\theta_2)\bigl(\frac{e}{ \theta_1+2\pi n}
+g\cot(\frac{g}{e}(\theta_2+2\pi m))+g\bigr)}\right.\displaystyle\frac{g}{ (\theta_1+2\pi
n)\tanh(\frac{g}{e}(\theta_2+2\pi
m))}\,-\\
&-& \left.\displaystyle\sum_{n,m=1} \frac{\mathrm { e } ^
{-S^{-}_{n,m}}\cos^2(\frac{\theta_1 -\theta_2 } { 2 } ) }
{\sin(\theta_1+\theta_2)\bigl(\frac{e}{ 2\pi n-\theta_1}
+g\cot(\frac{g}{e}(2\pi m-\theta_2))+g\bigr)}\displaystyle\frac{g}{ (2\pi n
-\theta_1)\tanh(\frac{g}{e}(2\pi m - \theta_2))}\right\},
\end{array}
\end{equation}
\end{widetext}
with

\begin{equation}\nonumber
\begin{array}{lll}
S_{n,m}^{+}&=&\frac{m_e^2}{eE}\,\theta_1^{+(n)}+\frac{M_d^2}
{eE}\,\theta_2^{+(m)}-\frac{m_eM_d}{eE}\sin(\theta_1
+\theta_2), \vspace{0.2cm}\\
S_{n,m}^{-}&=&\frac{m_e^2}{eE}\,\theta_1^{-(n)}+\frac{M_d^2} { eE
}\,\theta_2^{-(m)}+\frac{m_eM_d}{eE}\sin(\theta_1 +\theta_2).
\end{array}
\end{equation}
This sum looks rather ugly, however, the contributions of higher
winding paths are suppressed by the factor of
$\exp[-(\frac{m_e^2}{eE}2\pi n+\frac{M_d^2}{eE}2\pi m)]$. So, for
practical calculations only the leading term should be left in the
sum. The leading term is the one with ``$+$'' and zero winding
numbers. It is given by
\begin{widetext}
\begin{equation}
\mathrm{Im}\,\delta
M_m=-\frac{\lambda^2}{4\sqrt{2}\pi}\frac{eE}{M_m}\,
\mathrm{e}^{-S_0}\frac{ \cos^2(\frac{\theta_1
-\theta_2}{2})}
{\sin(\theta_1+\theta_2)\bigl(\frac{e}{
\theta_1}
+g\cot(\frac{g}{e}\theta_2)+g\bigr)}\frac{g}{\theta_1
\tanh(\frac{g}{e}\theta_2)},
\end{equation}
\end{widetext}
with the corresponding value of $S_0$
\begin{equation}\nonumber
S_0=\frac{m_e^2}{eE}\theta_1+\frac{M_d^2}{eE}
\theta_2-\frac{m_eM_d}{eE}\sin(\theta_1+\theta_2).
\end{equation}

\section{\label{thirmod} Bound state in 2D}
 If previous
considerations are reduced to two dimensions, then the situation
would be technically simpler, because instead of a monopole one
would have a free scalar particle, and a fermion--antifermion pair
instead of a dyon and a charged fermion. Thus the problem studied
above directly reduces to the decay of a bound state into a
fermion--antifermion pair in Thirring model. For an induced
Schwinger process in Thirring model there exists a calculation of
the pre-exponential factor in terms of the dual (Sine-Gordon) theory  by Gorsky and
Voloshin~\cite{Gorsky:2005yq}. This process is forbidden, as the
bound state is lighter than the two fermions, but again, it becomes
allowed when an external field is on.

One should note here that the first bound state of the massive
Thirring model should rather be rendered as a pseudoscalar. Due to
duality, a bound state in Thirring model corresponds to a special
kind of a soliton-antisoliton classical configuration (the so-called
``doublet'') in the Sine-Gordon model. A fermionic current $j^\mu$
corresponds in the dual picture to the topological current in
sine-Gordon
$$\bar{\psi}\gamma^\mu\psi=\epsilon^{\mu\nu}\partial_\nu\phi$$
which can be rewritten as
$$\bar{\psi}\sigma^3\gamma^\mu\psi=
\partial^\mu\phi.$$

This suggests that the matrix element $\langle
0|\bar{\psi}\sigma^3\psi| \pi\rangle$ is non-zero, $\sigma^3$
playing the same role for the 2-dimensional case as $\gamma^5$ for
the 4-dimensional. Thus an ``effective vertex'' for the considered
2D case should necessarily contain the $\sigma^3=-i\sigma^1\sigma^2$
Pauli matrix. Let us show the final result of the calculation. Here
the resummation over winding numbers is done exactly, factors like
$\frac{1}{1-e^{-\frac{2\pi \mu^2}{eE}}}$ being a consequence
thereof, $\mu_1$ and $\mu_2$ denoting masses of the fermions, which
are held arbitrary for the sake of generality:
\begin{equation}\nonumber
\begin{array}{rcl}
\displaystyle\mathrm{Im}\,\delta
m&=&\frac{-\lambda^2}{4m
\left(1-e^{-\frac{2\pi\mu_1^2}{eE}}\right)\left(1-e^{-\frac{ 2\pi
\mu_2^2}{eE}}\right) \sin(\theta_1+\theta_2)}\times \\
&\times&
\left\{\mathrm{e}^{-S^+_0}
\left[2\cos^2\left(\frac{\theta_1-\theta_2}{2}
\right)-\frac{eE}{\mu_1\mu_2}
\frac{1}{\sin(\theta_1+\theta_2)}\right]\right.-\vspace{0.2cm}\\
&&-\left.
\mathrm{e}^{-S^-_0}
\left[2\cos^2\left(\frac{\theta_1-\theta_2}{2}
\right)+\frac{eE}{\mu_1\mu_2}
\frac{1}{\sin(\theta_1+\theta_2)}\right]

\right\},\end{array}
\end{equation}

where
\begin{equation}\nonumber
\begin{array}{c}
\theta_{1}=\cos^{-1}\frac{m^2+\mu_1^2-\mu_2^2}{ 2m\mu_{1}}\\
\theta_{2}=\cos^{-1}\frac{m^2-\mu_1^2+\mu_2^2}{
2m\mu_{2}}\end{array}
\end{equation}

\begin{equation}\nonumber
\begin{array}{c}
S^{+}=\frac{\mu_1^2}{eE}\theta_1+\frac{\mu_2^2}
{ eE
}\theta_2-\frac{\mu_1\mu_2}{eE}\sin(\theta_1
+\theta_2),\\
S^-=\frac{\mu_1^2}{eE}(2\pi -\theta_1)+\frac{\mu_2^2}
{ eE
}(2\pi -\theta_2)+\frac{\mu_1\mu_2}{eE}\sin(\theta_1
+\theta_2)
\end{array}
\end{equation}

Note that $\lambda$ is an essential parameter here, having the
dimension of a mass. Thirring model calculations for a decay of
bound state with mass $m$ into two fermions with equal masses $\mu$
lead us to
\begin{equation}\nonumber
\mathrm{Im}\,\delta
m=-\frac{\lambda^2}{4m}\frac{\mathrm{e}^{
-S_0}}{\sin2\theta}
\left(2-\frac{eE}{\mu^2}
\frac{1}{\sin 2\theta}\right),
\end{equation}
where
\begin{equation}\nonumber
\theta=\cos^{-1}\frac{m}{2\mu}
\end{equation}
(resummation factor  $\frac{1}{\left(1-e^{-\frac{2\pi
\mu^2}{eE}}\right)^2}$ omitted here).

On the other hand, the decay rate in Thirring model in the strong
coupling limit (weak coupling limit of Sine-Gordon model) is
given~\cite{Gorsky:2005yq} as

\beq\nonumber \Gamma=\frac{4g\mu}{\pi^3}e^{-S_0} \eeq where
$g$ is Thirring
coupling constant, $g\gg 1$; $\mu$ is the mass of Thirring fermions,
$S_0$ is the classical action. Let us suggest that the
external meson
is the lightest bound state in the theory, for which in the
mentioned limit $m=\frac{\pi^2\mu}{2g}$. It has been obtained by us
$\Gamma=2\mathrm{Im}\delta m=\frac{4\lambda^2
g^2}{\mu\pi^4}e^{-S_0}$ in terms of Thirring model parameters.
Comparison of these two formulae yields

\beq\nonumber \lambda=\mu\sqrt{\frac{\pi}{g}} \eeq which
restores coupling
constant  $\lambda$ {\it in an induced Schwinger process} for the
lightest Thirring meson.

\section{\label{final} Conclusion}
The pre-exponential subleading asymptotic is obtained  for the
non-perturbative monopole decay into a charged fermion and a dyon
in 3+1 dimensions, as well as for the decay of a bound state into
a fermion-antifermion pair in 1+1 dimensions. These are the main
features of our work, since these quantities have never been
estimated before up to this order. In the two-dimensional case the
``effective vertex'' $\lambda\sim\frac{\mu}{\sqrt{g}}$ has been
restored for the decay of a bound state in the Thirring model.
Generalization to inhomogeneous fields, thermal field theory as
well as to charged fermion decay into a monopole-dyon pair are
going to be considered as the next problems.

\begin{acknowledgments}
Authors are indebted to A.~S.~Gorsky for suggesting this problem and
fruitful discussions, to E.~T.~Akhmedov,  F.~V.~Gubarev,
A.~V.~Mironov and A.~Yu.~Morozov for their useful comments. One of
us (A.Z.) would like to thank D.~V.~Shirkov for his friendly advice
and moral support. This work is supported in part by RFBR
04-02-17227 (A.Z.); RFBR 04-01-00646 Grants and NSh-8065.2006.2
grant (A.M.).
\end{acknowledgments}

\section{Appendix}
In order to obtain the (\ref{G_0_E}) one can act in
the following way. Green function of an electrically and
magnetically charged particle can be represented in terms of
a Feynman path integral:
\begin{equation}
\begin{array}{lll}
\label{matrix element} \langle
y|\frac{1}{m^2-D^2+(eF_{\mu\nu}+g\tilde{F}_{\mu\nu})\sigma^{\mu\nu}}|x\rangle=\mathrm{e}^{\frac{1}{2}(eEs\gamma^0\gamma^3
+igEs\gamma^1\gamma^2)}\times
\\\times\int\mathcal{D}x_{\|}\mathrm{e}^{-\int_0^s(\frac{\dot{x}_\|^2}{4}+ieA_\|\dot{x}_\|)dt}
\int\mathcal{D}x_{\perp}\mathrm{e}^{-\int_0^s(\frac{\dot{x}_\perp^2}{4}+igA_\perp\dot{x}_\perp)dt}.
\end{array}
\end{equation}
The above integrals are Gaussian, so the Green function can be
calculated by means of steepest descent method. The value of the
on-shell action is given in (\ref{action_E}). The preexponential
factor is given by a product of two determinants for $\|$ and
$\perp$ components being proportional to
\begin{equation}
\begin{array}{lll}
\frac{1}{\sqrt{\det(\|)}}\sim\frac{eE}{\sin\frac{eEs}{2}}, \\
\frac{1}{\sqrt{\det(\perp)}}\sim\frac{gE}{\sinh\frac{gEs}{2}}.
\end{array}
\end{equation}
Collecting everything together one gets expression
(\ref{G_0_E}).\\
The differential operator $m-\hat{D}_y$ (Euclidean version of
(\ref{Dirac operator})) acts only on the terms that contain variable
$y$. This action gives the values of $a_\|$ and $a_\perp$
\begin{equation}
\begin{array}{lll}
&\Bigl(m-\gamma^\mu\bigl(\partial_\mu+ieA_\mu(y)+ig\tilde{A}_\mu(y)\bigr)\Bigr)\mathrm{e}^{-S_s(y,x)}&=
\\ =&\Bigl(m+\frac{eE}{2}\gamma^\|(y-x)_\|\cot\frac{eEs}{2}+ \\
+& \frac{eE}{2}\gamma^0(y_3-x_3)-\frac{eE}{2}\gamma^3(y_0-x_0)+\\
+&\frac{gE}{2}\gamma^\perp(y-x)_\perp\coth\frac{gEs}{2}+ \\
+& i\frac{gE}{2}\gamma^1(y_2-x_2)-i\frac{gE}{2}\gamma^2(y_1-x_1)\Bigr)\mathrm{e}^{-S_s(y,x)}=\\
=&\bigl(m+\gamma_\mu a^\mu(y,x)\bigr)\mathrm{e}^{-S_s(y,x)}.
\end{array}
\end{equation}
Formula (\ref{Trace}) can be obtained by substituting the
propagator in the expression
(\ref{contr}) by  (\ref{fermion propagator})
\begin{equation}
\mathrm{tr}(G_eG_d)=\mathrm{tr}\Bigl((M_d+a^d_\mu\gamma^\mu)G_{E}^{d(0)}(m+a^e_\mu\gamma^\mu)G_{E}^{e(0)}\Bigr).
\end{equation}
Note that $G^{(0)}_E$ also possesses matrix structure (see
(\ref{matrix element})).
Using the well-known formula for the exponent of a combination of
$\gamma$-matrices one gets (\ref{calculated trace}). After the trace
has been calculated,  integration over $w$ and $z$ should be
done, being of the form
\begin{equation}
\begin{array}{lll}
\displaystyle\int d^2z_\|d^2w_\|d^2z_\perp d^2w_\perp(B+A_\|(w-z)_\|^2+A_\perp(w-z)^2_\perp)\times\\
\displaystyle\times\displaystyle\mathrm{e}^{\displaystyle-(a_\|(w-z)^2_\|+b_\|z_\|^2+c_\|(y-w)^2_\|)}\times
\\
\times\displaystyle\mathrm{e}^{\displaystyle-(a_\perp(w-z)^2_\perp+b_\perp
z_\perp^2+c_\perp(y-w)^2_\perp)-2\epsilon(w_1z_2-w_2z_1)}= \\
=(B-A_\|\frac{\partial}{\partial a_\|}-
A_\perp\frac{\partial}{\partial
a_\perp})\frac{\pi^4\mathrm{e}^{-\frac{T^2}{\frac{1}{a_\|}+\frac{1}{b_\|}+\frac{1}{c_\|}}}}
{(a_\|b_\|+a_\|c_\|+b_\|c_\|)(a_\perp b_\perp+a_\perp
c_\perp+b_\perp c_\perp)},
\end{array}
\end{equation}
where $a$, $b$, $c$, $A$, $B$, $\epsilon$ are some
constants. If the values of these constants are
substituted one obtains (\ref{delta G
befor int A}).

\end{document}